\documentclass[twocolumn]{openjournal}

\renewcommand{\doi}[1]{\href{https://doi.org/#1}{DOI: \nolinkurl{#1}}}

\usepackage{xcolor}
\usepackage{textgreek}
\usepackage[utf8]{inputenc}
\usepackage[english]{babel}
\usepackage{amsmath}
\usepackage{listings}
\definecolor{backcolour}{rgb}{0.9, 0.95, 0.93}

\usepackage{hyperref}
\hypersetup{
    unicode, 
    colorlinks=true,
    linkcolor=linkcolor,
    citecolor=linkcolor,
    filecolor=linkcolor,
    urlcolor=linkcolor,
}
\usepackage{color,colortbl}
\definecolor{linkcolor}{rgb}{0.0,0.3,0.5}
\usepackage{tensind}
\tensordelimiter{?}
\DeclareGraphicsExtensions{.bmp,.png,.jpg,.pdf}
\usepackage{verbatim}
\usepackage[normalem]{ulem}
\usepackage{orcidlink}
\usepackage{soul}
\graphicspath{ {./figures/} }

\begin{document}
\title{Fast and accurate bandpass integration of complex SEDs with neural networks}

\author{L.~A.~Bianchi\orcidlink{0009-0002-6351-5426}}
\email{l.a.bianchi@astro.uio.no}
\affiliation{Institute of Theoretical Astrophysics, University of Oslo, P.O. Box 1029 Blindern, N-0315 Oslo, Norway}

\begin{abstract}
    The paper presents a novel ML-based approach for fast and accurate bandpass integration of spectral energy distributions (SEDs).
    The non-locality of the integral operator involved guarantees the smoothness of the target function to be approximated, making it an excellent use-case for neural networks.
    The computational method developed for this work has been wrapped within \texttt{pyfine} (\textit{FINE}: Fast Function Interpolation via NEural NEtworks). This new Python package makes the relevant code available and easy to use by everyone within the scientific community.
    The method is demonstrated with two different bandpass integration test cases with 3 and 9 free parameters, respectively, where both accuracy and computational performance are analyzed.
    The results show that the method is capable of predicting values of the bandpass integral with maximum relative residuals as low as 0.002\% and 0.08\% respectively, within the sampled region of the parameter space, fast run times, and very little memory overhead.
    The usage of \texttt{pyfine} is especially effective when the computation of an SED functional form, given a set of values for its free parameters, is expensive.
    However, already in the first of the two test cases, where a simple modified black body SED was employed, the neural network's measured run times resulted faster by two orders of magnitudes than an optimized multithreaded version of the ``brute force'' exact calculation of the integral.
\end{abstract}

\begin{keywords}
    {First, second}
\end{keywords}

\maketitle

\section{Introduction}
\label{sec:intro}
When observing the sky, an astronomical detector typically collects the incoming radiation over a finite range of frequencies.
An ideal instrument, with a perfect top hat-shaped response, would weigh all of the incoming radiation equally, within a nominal frequency interval, and completely block any power outside it.
However, in an actual instrument, due to real-world non-idealities, the contribution of each single frequency ``bin'' is weighted differently than the others, and the two ends of the collected frequency range are never sharp cut-offs as in ideal designs.
The resulting profile describing the instrumental response in the frequency domain is typically referred to as \textit{bandpass}, or responsivity, and it is a fundamental piece of the instrumental understanding, as it directly relates the signal recorded to what we are ultimately interested in, the real observed sky.

In practice, bandpass corrections are carried out as an integral operation, wherein a specific sky signal, modeled as a spectral energy distribution (SED) is weighted by the instrumental responsivity, and integrated over the relevant frequency range. 
Within astrophysical data analysis, growing importance is being given to evaluating the bandpass corrections accurately when calibrating observations, as the instrumental sensitivity, especially of space-based missions, keeps improving~\citep{Hoang_2017, Ward_2018, Ghigna_2020, Giardiello_2024}.

There are MCMC-based analysis methods, such as Commander~\citep{Eriksen_2004}, which provide a detailed statistical exploration of the interplay between instrumental effects and astrophysical emission components through iterative sampling techniques, such as Bayesian Gibbs sampling.
In this setting, a precise estimate of the sky model is needed to correctly calibrate the instrumental response and an accurate instrumental characterization, including the bandpass, is needed to accurately solve for the true sky emission.
Such methods are very effective in breaking the degeneracies within the astrophysical and instrumental analysis, e.g.~\citep{watts2023_dr1}, but are typically very computationally expensive, as thousands of iterations are required for a complete exploration of the parameters' posterior distributions.
Each step of the pipeline ends up being executed over and over, making the run time performance a key aspect of each operation.
The bandpass integral is no exception to this, and computational methods for evaluating it quickly and with a low error rates are required to obtain feasible run times. 
This is especially relevant for full-sky and high-resolution analysis, where a new bandpass correction must be recomputed for each of the millions (or even billions) of pixels of the processed maps. 

There is a second aspect driving the motivation for fast and accurate bandpass integration methods: the complexity of the SED modelling the source signal.
While originally most astrophysical sources of diffuse radiation where generally described by simple models such as modified black bodies or power laws, as the physical comprehension of the Galactic foregrounds advances, more accurate, and computationally expensive, models are being published.
Astrodust+PAH~\citep{Hensley_2023}, for example, provides a state-of-the-art description of the interstellar dust emission, through a complex SED featuring more than 10 parameters, and run times for its computation which are incompatible with high-resolution MCMC iterative analysis.
To overcome the slow computation of bandpass integrals, precomputation and interpolation methods such as gridding and splining have traditionally been implemented within pipelines such as Commander.
However, none of those can really handle a number of parameters of 3 or more, making them unable to target complex SEDs integration, such as Astrodust+PAH.

The solution proposed by this work, as described in the next section, is to employ a neural network to efficiently predict the value of the bandpass integral, directly from the parameters describing the SED. 
Sections~\ref{sec:examp1} and~\ref{sec:examp2} provide two examples of application of the method presented, to a simpler modified-black-body case and a more complex one based on Astrodust+PAH respectively.
Sect.~\ref{sec:pyfine} gives an overview of a new Python package created to wrap the computational tools developed for this work.

\section{Method}\label{sec:method}
\subsection{Bandpass integral}
In many astronomical experiments, the bandpass plays a fundamental role as it relates the instrument's physical response to the underlying sky signal. 

In fact, since no detector is perfectly sensitive to just a single frequency, the effective sky signal observed by a radiometer at any given time is determined by the integral
\begin{equation}
    I(\mathbf{\Theta}) = \int s(\nu, \mathbf{\Theta})\, \tau(\nu)\,d\nu,
\label{eq:integ}
\end{equation}
where $s(\nu, \mathbf{\Theta})$ is the true underlying sky signal of the given source, modeled through an SED whose functional form generally depends on a series of parameters (here grouped under the notation
$\mathbf{\Theta}$), and $\tau(\nu)$ represents the bandpass of the instrument, which gives a measure of the portion of the radiation that gets collected by the detector, at each frequency $\nu$.

The bandpass profile is typically measured in a laboratory by the manufacturer of the instrument itself, and as such should in theory be a well known fixed quantity. 
In practice, also the bandpass can depend on a set of unknown parameters that must be fit during the analysis, e.g.~\citep{bp10}.
For simplicity, such effects are not being considered throughout the current section.

In practice, once a functional form of the SED is given, solving Eq.~\eqref{eq:integ} only reduces to a sum over an array of points, typically few thousands, describing the bandpass profile, multiplied by the SED evaluated at those corresponding frequencies. 
However, when performing a full-sky analysis, or a wide portion of it, it is often necessary to allow the physical parameters $\mathbf{\Theta}$ to vary across the sky, and with them the resulting radiation described by the corresponding SED.
In this setting, not only the summation must be repeated for each pixel of the sky map, but the computation of a new SED, which is a function of $\nu$, is required as well. 
Such operations, especially for SEDs described by complex functional forms, can lead to prohibitive run times if repeated over the millions of pixels composing high resolution sky maps.

\subsection{Neural Network approach}\label{ssec:NNapproach}
The approach proposed in this work is based on training a feed-forward neural network (FFNN, for short) for each bandpass-SED combination, where the parameters $\mathbf{\Theta}$ constitute the input of the network and the corresponding integral value $I(\mathbf{\Theta})$ is the target.
The idea is to mold the network directly on a specific bandpass shape and parametric SED form through the training, bypassing the integral computation entirely.
Reaching fast inference times, while preserving a relative error on the integral estimate as low as possible\footnote{Within Commander framework, for example, the maximum acceptable error on the calculated integral value is generally set at 0.1\%.}, is in fact the main goal of this work.

There has been mainly two points suggesting the potential success of a neural networks-based approach; (i) given the non-locality of the integral operator, the target function will naturally be smooth and differentiable anywhere in the parameter space, without any abrupt change in magnitude, offering a perfect use case for small neural networks.
(ii) Training data is extremely easy to produce, and virtually unlimited in size. 
A dataset simply consist in a set of random realization of the parameters, within physically reasonable boundaries, paired with the corresponding true integral values, computed through traditional ``brute-force'' summation, as targets (or \textit{labels}). 

\begin{figure}[]
    \centering
    \includegraphics[width=.95\linewidth]{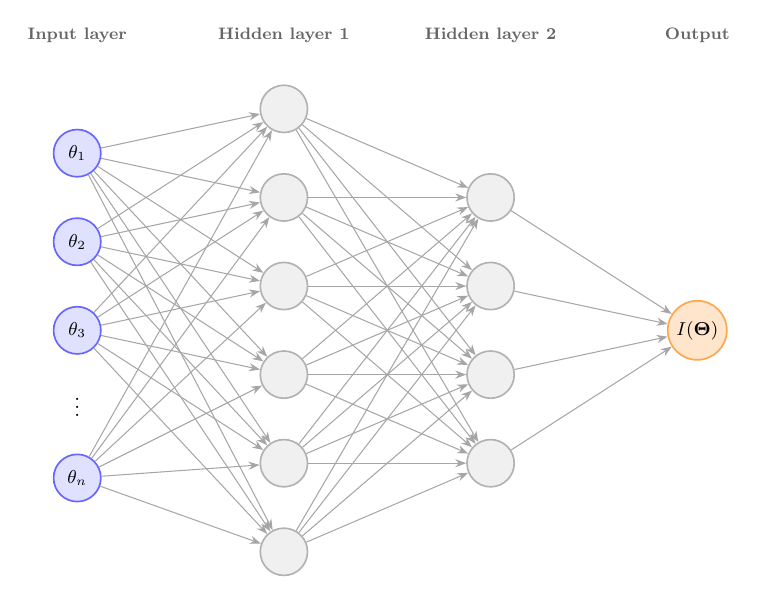}
    \caption{Diagram of typical neural network designed for the purpose of this work. Here the parameter array is given by $\mathbf{\Theta} = (\theta_1, \theta_2, \theta_3, \dots, \theta_n)$ which corresponds to the nodes of the input layer of the network.}
    \label{fig:nn}
    \vspace{8.pt}
\end{figure}

A typical neural network required for this task is necessarily designed with as many input nodes as free parameters and a single output node yielding the result of the integral evaluation. 
Fig.~\ref{fig:nn} displays an example of such architecture.
The number and shape of hidden layers strongly depends on the number of input parameters and the complexity of the integral function.
However, the two most common configurations are typically a pyramidal shape, where the first hidden layer is several times wider than the input layer, and the subsequent ones are progressively narrower, generally by a factor of 2, and a ``brick'' shape, where identical hidden layers follow each other.


\begin{figure}[]
   \centering
   \includegraphics[width=.95\linewidth]{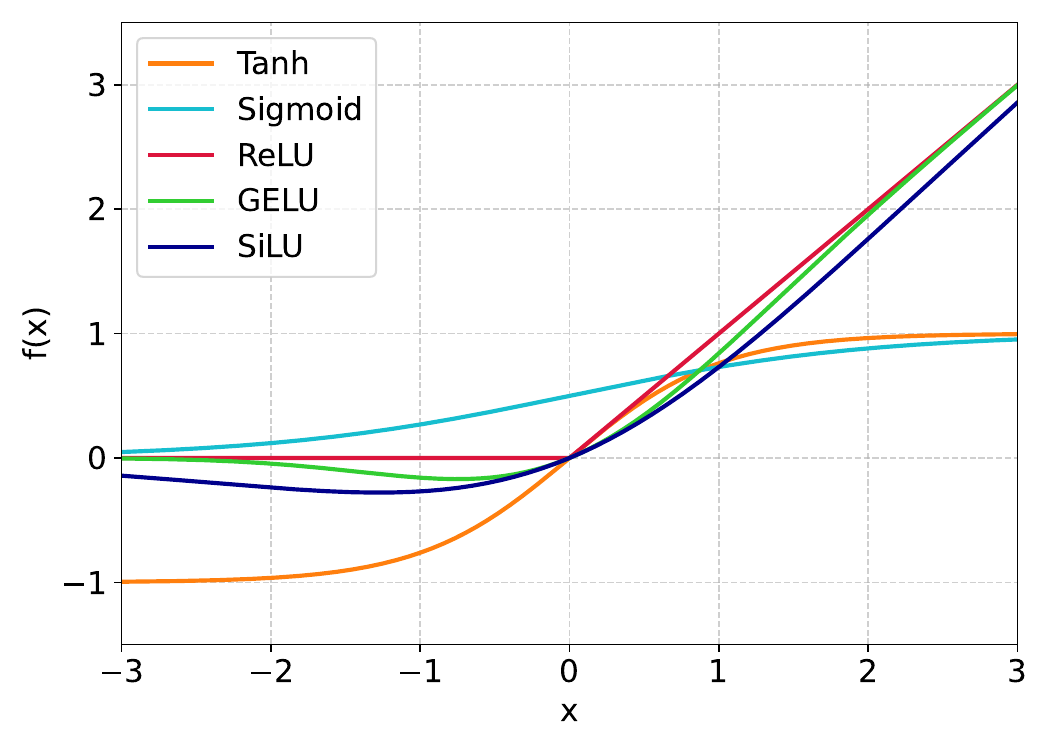}
   \caption{Comparison of some of the most popular activation functions in modern deep learning.}
   \label{fig:activations}
   \vspace{10.pt}
\end{figure}


\subsection{Activation function}\label{ssec:af}

When designing a neural network for a specific problem, an important choice resides in the activation function coupled to the hidden layers.
The role of the activation function is mainly two-fold: it introduces non-linearity throughout the network, which is necessary for it to learn the details of the target function hyper-surface, and allows to control the magnitude of the signal propagated between the layers, helping to stabilize the training.
For a clear and comprehensive introduction to the role of the activation functions in deep learning, see chapter 6 of~\citep{Goodfellow-et-al-2016}.

Within modern machine learning there are many possible activation functions to choose from, e.g., ReLU~\citep{ReLU}, GELU~\citep{GELU}, tanh or sigmoid, shown in Fig.~\ref{fig:activations}.
Giving a comprehensive description and comparison between each of them is beyond the scope of this work, for an in-depth analysis of different activation functions and their performances see~\citep{activfuncs}.

For small neural networks aiming for fast inference time, the default choice is usually the ReLU for its simplicity and extremely low computational cost. Its definition in fact simply reads
\begin{equation}
    \text{ReLU}(x) = \max(0, x) = \begin{cases} x & \text{if } x > 0 \\ 0 & \text{otherwise} \end{cases}
\end{equation}
However, as shown in figure~\ref{fig:activations}, the sharp knee centered on zero makes it unideal to approximate smooth functions as the ones targeted here.
The smoothness of the target function has in fact been the main reason for choosing the Sigmoid Linear Unit~\citep{SiLU}, or SiLU, as default activation function for this work.
It has been proven to be particularly effective on different kinds of deep learning tasks, while maintaining a lower computational cost than more complex functions such as the GELU.
It is mathematically defined as
\begin{equation}
    \text{SiLU}(x) = x \cdot \sigma(x) = x \cdot \left( \frac{1}{1 + e^{-x}} \right) = \frac{x}{1 + e^{-x}},
\end{equation}
where $\sigma(x)$ is the sigmoid function. 
The SiLU combines the smoothness and differentiability of the sigmoid or tanh functions, as seen in Fig.~\ref{fig:activations} with the effectiveness against saturation and vanishing gradients of the ReLU.

\begin{figure}[]
    \centering
    \includegraphics[width=.95\linewidth]{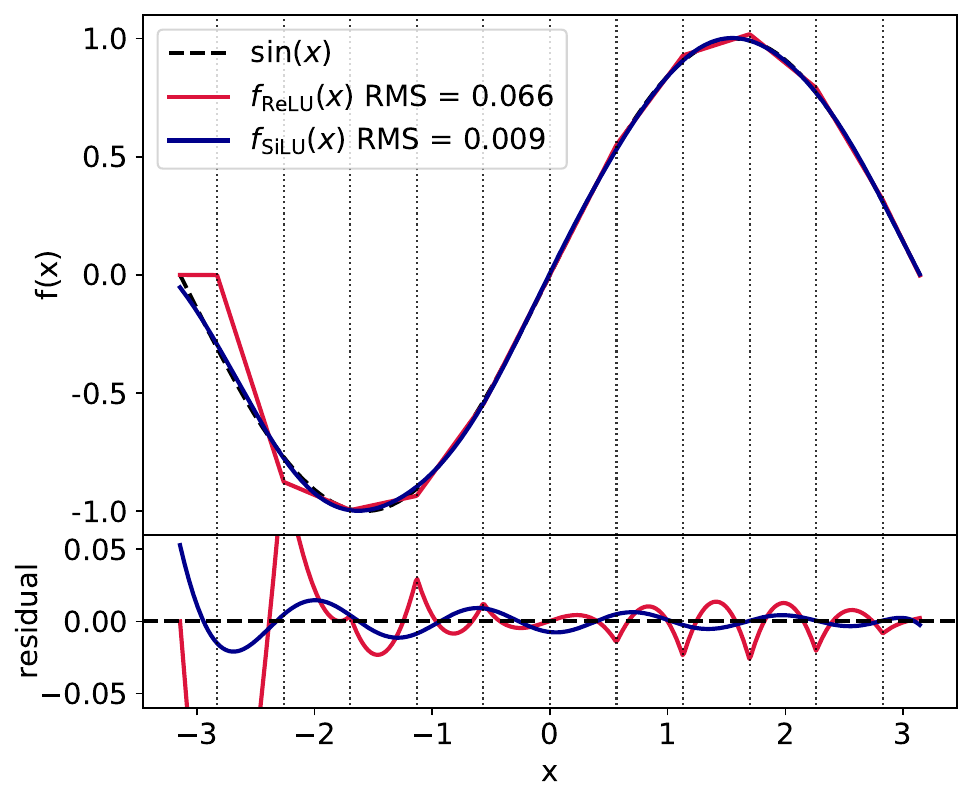}
    \caption{The plot shows the result of approximating a smooth sine function through the regression defined in~\eqref{eq:radial_basis}, both using a ReLU or SiLU as basis function. The vertical dotted lines show the position of the centers $c_i$.}
    \label{fig:relu_vs_silu_1d}
    \vspace{10.pt}
\end{figure}

To give a better idea of the impact of the activation function and why the SiLU is expected to deliver better accuracy than the ReLU, the following 1-d example is provided.
A sine curve is approximated through a fixed number of ReLUs and SiLUs, used as basis functions pinned around a fixed number of centers, which are mocking the layers of the neural network, in a sort of radial basis function regression;
\begin{equation}\label{eq:radial_basis}
    \sin(x) \approx f_\phi(x) := \sum_{i=1}^{N}\,w_i\,\phi(x - c_i),
\end{equation}
where the $N = 11$ centers $c_i$ are equispaced over the interval $[-\pi, \pi]$, and the weights $w_i$ are determined through a simple least-squares optimization.
The comparison between the two cases, $\phi(x) = \mathrm{ReLU}(x)$ and $\phi(x) = \mathrm{SiLU}(x)$, is shown in Fig.~\ref{fig:relu_vs_silu_1d}.
While a neural network is of course a far more complex case, and of higher dimensionality, this example shows the intuition behind why a smoother activation function is expected to provide higher accuracy given a fixed network depth. 

\subsection{Training}\label{ssec:method-training}
The supervised training of a feed-forward neural network is a theoretically robust and straightforward process composed by two main steps repeated one after the other for a number of so-called \textit{epochs}. \textbf{(i) Forward evaluation}: the network is used in the current state to predict a value for the target, which is quantitatively compared to the ground truth given by the training dataset through a loss (or cost) function. \textbf{(ii) Backward propagation}: the derivative chain rule is applied backwards through the network to update the weights of each neuron according to the gradient of the loss function.

In practice, however, training a neural network is a complex procedure where many parameters and components can be adjusted, such as which loss function to use, the size of the training data batches, or the algorithm used for optimizing the weights.
To give a comprehensive description of all the concepts and quantities involved, and their interplay during training, is out of the scope of this work, and we will here only outline the main ones.
A great source of information on this regard can again be found in~\citep{Goodfellow-et-al-2016}.

A preliminary phase of research and test, wherein many models and training parameters have been tried, helped to determine a particularly effective and flexible training setup that has been kept more or less untouched through the work here presented.
The main features are:
\begin{itemize}
    \item Adaptive Moment Estimation, or Adam~\citep{Adam}, as optimizer algorithm. It is one of the most widely used in the entire field of machine learning for its efficiency and robustness, as it is able to dynamically adjust the learning rate for each of the trained parameters.
    \item Huber~\citep{huber_loss} loss. Defined as 
        \begin{equation}\label{eq:Huber}
            L_{\delta}(r) =
            \begin{cases}
                \dfrac{r^{2}}{2}  & \text{if } |r| \leq \delta \\[10pt]
                \delta \left(|r| - \dfrac{\delta}{2}\right) & \text{otherwise}
            \end{cases},
        \end{equation} 
        where $r$ is the residual between the predicted target and ground truth, and $\delta$ is an arbitrary parameter typically set to 1, it behaves as a mean absolute error (MAE) for large residuals, and as a mean squared error (MSE) cost function for small ones. It has the peculiarity of being very robust and stable against outliers.
    \item The usage of a ``reduce on plateau'' learning rate scheduler algorithm, which keeps track of the loss function throughout training and cuts down the learning rate, typically by halving it, when a flattening trend is detected, allowing for a deeper exploration of that minima. The learning rate, in fact, gives a measure of the step size taken by the search algorithm when exploring the landscape of the trainable parameters.
    \item An ``early-stopping'' routine is useful to keep track of the training trend, and interrupt it when no significant improvement has been detected.
    \item Both input and targets are normalized according to
    \begin{equation}
        x_{\mathrm{norm}} = \frac{x - \hat{x}}{\sigma_x},
    \end{equation}
    where $\hat{x}$ and $\sigma_x$ are the sample mean and standard deviation of a general component $x$ of the training dataset.
    Normalizing the inputs improves the gradient symmetry and smoothness of the loss landscape.
    Normalized targets help keeping the overall magnitude of the gradients low and avoid overshooting during the search, improving training efficiency~\citep{lecun1998efficient}.
\end{itemize}
This training setup aims to reach the best possible performance for any kind of network employed with minimal tuning required from the user.
There are however a few parameters which are naturally dependent on the network size, such as the training batch size, or the initial learning rate, which might have to be tuned on the specific use case for an optimal result.

\section{Implementation: \texttt{pyfine} Python package}\label{sec:pyfine}
\subsection{Introduction}
As described in Subsect.~\ref{ssec:NNapproach}, and shown in Sections~\ref{sec:examp1} and~\ref{sec:examp2}, the task covered by the neural networks employed is ultimately to approximate, with high accuracy and short inference time, smooth functions of a certain set of parameters.
Such method can be useful for a variety of computationally-intensive data analysis applications which require frequent evaluations of smooth functions of 3 or more parameters, where standard gridding or splining methods start breaking due to high dimensionality.
For this reason, the code developed for this work has been wrapped into a new Python package, named \texttt{pyfine}\footnote{The package can be downloaded and installed as described at \url{https://github.com/LeeoBianchi/pyfine}.} (\textit{FINE}: Fast Function Interpolation via NEural NEtworks).
The main goal is to make the method presented easily applicable to any bandpass correction scenario, and possibly any smooth function approximation in general, reducing as much as possible the setup overhead for the user.

Currently, \texttt{pyfine} is mainly exposing a core class, \texttt{FFNN}, which wraps a general and flexible feed-forward neural network based on PyTorch~\citep{pytorch}, and an \texttt{utils} module, containing some other useful computational tools such as a training dataset generator.

\subsection{Usage}
The usage of \texttt{pyfine} is quite straightforward, and this section outlines the main steps to get started with it.
A new neural network can be initialized with a single line of code as
\begin{lstlisting}[basicstyle=\ttfamily\footnotesize, backgroundcolor=\color{backcolour}]
from pyfine.nn import FFNN

myNN = FFNN([3,16,8,4,1])
\end{lstlisting}
here the array [3,16,8,4,1] describes the shape of the network, where each element gives the number of nodes of the corresponding layer. 
The first number, i.e. the width of the input layer, must of course match the number of free parameters $\mathrm{n}_{\mathrm{params}}$ of the function being targeted.
A few other optional arguments are allowed, such as the activation function coupled to the hidden layers.

To generate a training dataset for applying the method to a scalar function \texttt{foo(a,b,c)}, it is convenient to use the specific \texttt{DatasetGenerator} class provided within \texttt{utils}, as shown in the brief example below.
\begin{lstlisting}[basicstyle=\ttfamily\footnotesize, backgroundcolor=\color{backcolour}]
from pyfine.utils import DatasetGenerator
from pyfine.utils import VectorScalarDataset

#define a function
def foo(a, b, c):
    return a**2 + 2*b**2 + 3*c**2 + 1

#sampling interval boundaries
lo = [-2, -1.5, -1]
hi = [ 2,  1.5,  1]

#number of samples
N = 100000

gen = DatasetGenerator(foo)
X, y = gen.generate_uniform(N, lo, hi)

dataset = VectorScalarDataset(X, y)
\end{lstlisting}
The two tensors obtained, \texttt{X} and \texttt{y}, are packed into a wrapper class, \texttt{VectorScalarDataset}, inherited from PyTorch's general dataset class, for a more convenient handling within the training routine.
To train the network on the dataset just created, it is enough to use the following few lines of code:
\begin{lstlisting}[basicstyle=\ttfamily\footnotesize, backgroundcolor=\color{backcolour}]
train_params = {
    "print_log"         : True,
    "max_epochs"        : 300,
    "train_batch_size"  : 512,
    "valid_batch_size"  : 4096,
    "best_model_outpath": 'my_NN.pth',
    }

myNN.fit(dataset, train_params)
\end{lstlisting}
where \texttt{train\_params} is an optional dictionary specifying a series of parameters (here only a few are shown) used to override the default values and customize the training routine.
Throughout training, the model with the best training loss will be kept track of, and automatically saved on disk at the path specified.
Later it can be loaded with a specific constructor method.
\begin{lstlisting}[basicstyle=\ttfamily\footnotesize, backgroundcolor=\color{backcolour}]
myNN = FFNN.from_file('my_NN.pth')
\end{lstlisting}
Lastly, the trained network can be used to infer the value of \texttt{foo} for entire batches of M samples at once, defined as a tensor \texttt{X\_m} of size (M,$\mathrm{n}_{\mathrm{params}}$).
\begin{lstlisting}[basicstyle=\ttfamily\footnotesize, backgroundcolor=\color{backcolour}]
M = 1000
X_m = torch.tensor(np.random.rand(M,3))
y_pred = myNN.predict(X_m)
\end{lstlisting}
The output \texttt{y\_pred} is naturally a tensor of shape (M,) containing the predictions.

\section{First test case: 3-parameters modified black body}\label{sec:examp1}
\subsection{Setup}
\begin{figure}[]
    \centering
    \includegraphics[width=.95\linewidth]{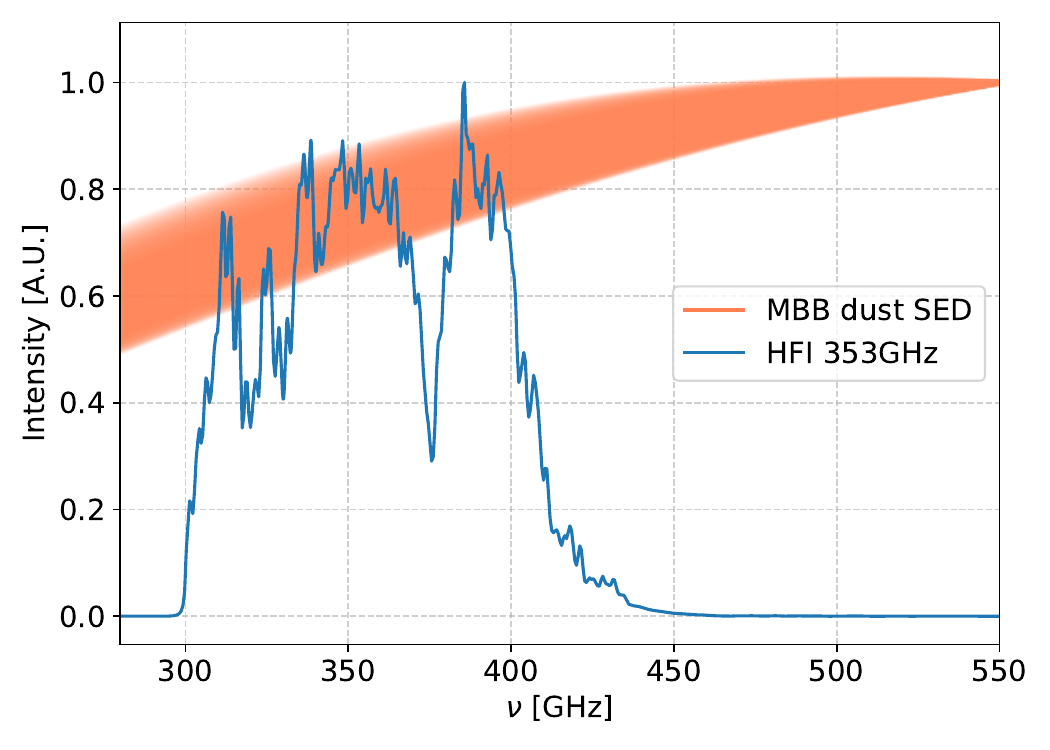}
    \caption{Panck HFI 353~GHz bandpass profile plotted together with a bundle of modified black body SEDs generated for a uniform grid of $\beta$ and $T$ within the explored region. Note that the y-axis is represented in arbitrary units, as multiplicative factors, such as unit conversions, will not affect the analysis presented.}
    \label{fig:case_1_bp}
    \vspace{10.pt}
\end{figure}

In this section we present a first simple application of the method, where a modified blackbody is taken as SED together with a bandpass from Planck 353\,GHz~\citep{HFI}.
While the main goal here is to build intuition about the scenario and possible challenges with a rather simple case, the modified black body is still very used for actual modeling of foreground emissions, e.g.~\citep{Planck2013_XI, Juvela_2024, CG02vi}.
In this setting, the emissivity is described by the following 2-parameter function of the frequency $\nu$:
\begin{equation}
    s(\nu; \mathbf{\Theta}) = s(\nu; \beta, T) = \left(\frac{\nu}{\nu_\mathrm{ref}}\right)^{\beta+1}\left(\frac{e^{h\nu_\mathrm{ref}/k_\mathrm{B}T}-1}{e^{h\nu/k_\mathrm{B}T}-1}\right).
\end{equation}
Where $\beta$ and $T$ are the free parameters and $\nu_\mathrm{ref}$ is a constant reference frequency of 545\,GHz.
The resulting SED curve is shown together with Planck 353 bandpass profile in Fig.~\ref{fig:case_1_bp}.

To add an extra bit of real-world complexity, a bandpass shift $\sigma$ is introduced, making the integral value $I(\mathbf{\Theta})$, defined in Eq.~\eqref{eq:integ}, dependent on the three parameters $\mathbf{\Theta} = (\beta,T,\sigma)$.
As mentioned in Sect.~\ref{sec:intro}, a 3-dimensional problem is already challenging for traditional interpolation methods, making this test case of potential interest for an actual analysis implementation.

\subsection{Training}\label{ssec:case1_training}
First, a training dataset is generated by drawing 1~M samples of $\beta$, $T$ and $\sigma$ from uniform distributions within the intervals $[1.75, 1.95]$, $[9.5, 12.5]\,\mathrm{K}$ and $[-10, 10]\,\mathrm{GHz}$ respectively, taken as physical ``prior'' boundaries for the parameters, plus roughly a 10\% extra coverage to ensure good interpolation accuracy on the region we are interested in.
The dataset is then completed by the targets, computed as the true integral value for each of the 1~M parameter tuples, giving a total size of around 30MB.
As a common procedure, the dataset is randomly split in two: the proper training dataset, and a validation dataset, with a relative size of 20\% of the former, only used for keeping track of the model's performance as the training runs, mainly to prevent any possible overfitting.

A selection of four neural networks (listed in Table~\ref{tab:MBB_4net}), taken as the four possible combinations between two different shapes, ``brick'' and pyramidal, and two activation functions, ReLU and SiLU is now analyzed. 

\begin{table}[ht]
    \setlength{\tabcolsep}{6pt} 
    \renewcommand{\arraystretch}{1.1} 
    \caption{Main features of the 4 tested neural networks}
    \centering
    \begin{tabular}{r | c c r}
        model \#  &  Shape  & Activation function &  $N_{\mathrm{t}}$ \\
        \hline
        1 & [3,12,8,1] & SiLU & 193 \\
        2 & [3,6,6,6,6,1] & SiLU & 157\\
        3 & [3,12,8,1] & ReLU & 193\\
        4 & [3,6,6,6,6,1] & ReLU & 157\\
    \end{tabular}
    \vspace{1mm}
    \begin{flushleft}
    \small
    \textit{Note --} The first column gives an arbitrary indexing, useful for referring to plots in Fig.~\ref{fig:training}. The network shape is given in the second column as an array, where each element gives the number of nodes of the corresponding layer. The shape [3,6,6,6,6,1] is referred to as ``brick'', while [3,12,8,1] as pyramidal. The last column gives the number of trainable parameters as a measure of computational complexity. The actual total number of FLOPs (floating-point operations) is harder to estimate, as complex activation functions as the SiLU should be included as well. However, a general rule of thumb is given by $N_{\mathrm{FLOPs}} \approx 2 \times N_{\mathrm{trainable}}$.
    \end{flushleft}
    \label{tab:MBB_4net}
\end{table}
The same training routine, outlined in Subsect.~\ref{ssec:method-training}, has been applied to all of the four models.
Given the small size of the networks employed for this use case, the training batch size has been fixed to 256, to give a good balance between training speed and parameter space exploration.

\begin{figure}[]
    \centering
    \includegraphics[width=.95\linewidth]{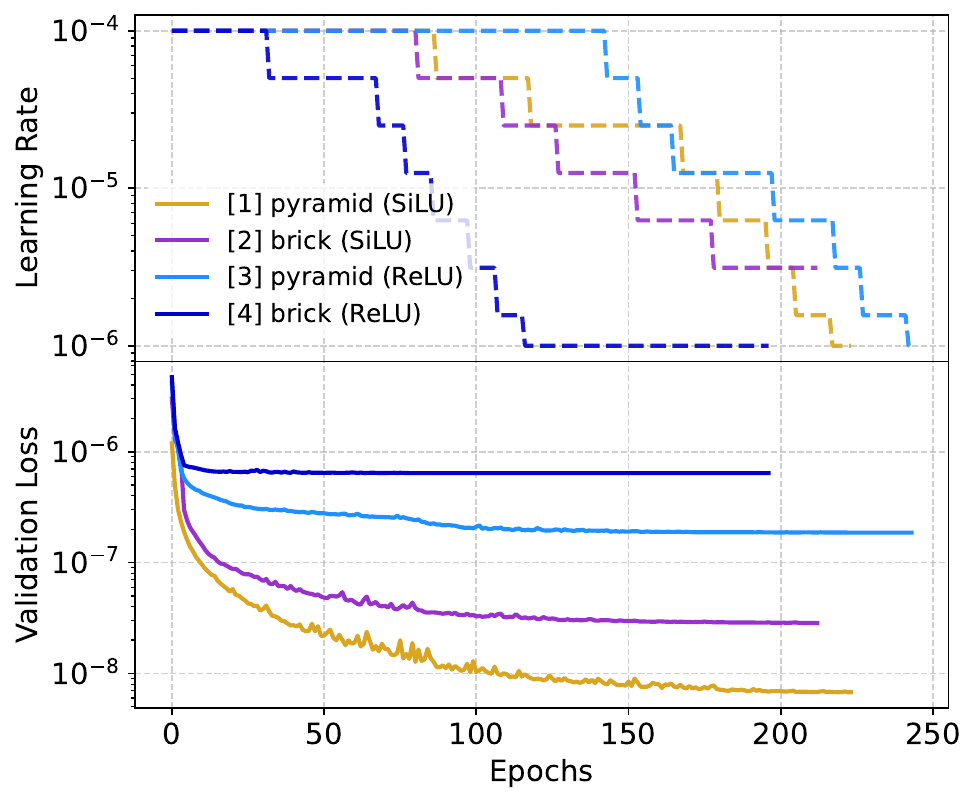}
    \caption{Training metrics of the four network architectures listed in Table~\ref{tab:MBB_4net}. The dashed lines in the top panel show the learning rate decay, dictated by the scheduler.
    The bottom panel shows the averaged validation loss function for each network, which is correlated to the residuals through Eq.~\eqref{eq:Huber}. }
    \label{fig:training}
    \vspace{10.pt}
\end{figure}

The plot shown in Fig.~\ref{fig:training}, which summarizes the training results of the four networks tested, provides a practical evidence of what has been described in Sect.~\ref{ssec:af}: given the same network shape, the SiLU activation function is providing a significantly better accuracy performance than the ReLU, with the validation loss function settling on values lower by 1 to 2 orders of magnitude.

\subsection{Accuracy}
\begin{figure*}[]
    \centering
    \includegraphics[width=0.8\linewidth]{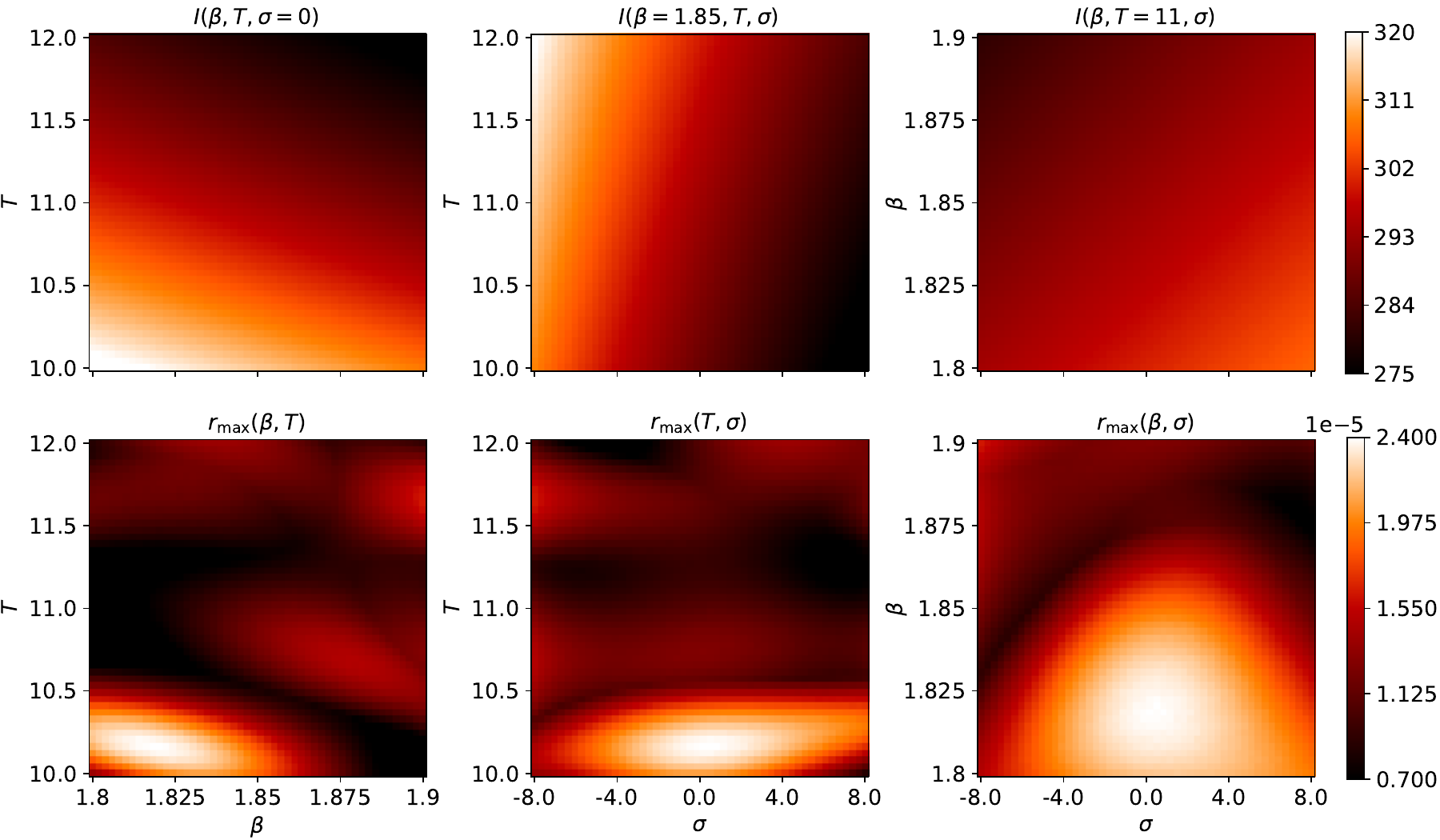}
    \caption{
    Inference accuracy analysis for the model \#1.
    The top row shows, from left to right, the integral value plotted for $\sigma$, $\beta$ and $T$ fixed one at a time respectively. The figures clearly show the smoothness of the integral function, as described~\ref{sec:method}.
    The bottom row shows, from left to right, the maximum, taken along the dimensions of $\sigma$, $\beta$ and $T$ respectively, of the relative residual as defined in~\eqref{eq:relres}. All the axis show the same range within which the parameters have been sampled for generating the test dataset.}
    \label{fig:valid_multifig}
    \vspace{12.pt}
\end{figure*}

To explore the accuracy of the networks trained in predicting the bandpass integral values, a new test dataset is produced by sampling the parameters on a $50 \times 50 \times 50$ regular grid defined over the intervals $[1.8, 1.9]$, $[10, 12]\,\mathrm{K}$ and $[-8, 8]\,\mathrm{GHz}$ for $\beta$, $T$ and $\sigma$ respectively. 
The corresponding exact integral values are also computed and then compared to the neural network predictions.
The relative residuals are computed across the sampled region of the parameter space as
\begin{equation}\label{eq:relres}
    r(\mathbf{\Theta}) = \frac{\left|I_\mathrm{true}(\mathbf{\Theta}) - I_\mathrm{pred}(\mathbf{\Theta}) \right| } {I_\mathrm{true}(\mathbf{\Theta})},
\end{equation}
where $I_\mathrm{true}(\mathbf{\Theta})$ and $I_\mathrm{pred}(\mathbf{\Theta})$ are the true and predicted values of the bandpass integral respectively.
Figure~\ref{fig:valid_multifig} shows, as an example, an analysis of the residuals in the sampled region of the parameter space for the model \#1 from Table~\ref{tab:MBB_4net}.
Table~\ref{tab:MBB_relres} is listing the median and maximum error measured on the whole test dataset for each of the four networks analyzed.
Again, the difference between ReLU and SiLU activation functions is well noticeable, with the latter providing maximum relative residuals around 20 times lower. 
\begin{table}[]
    \setlength{\tabcolsep}{6pt} 
    \renewcommand{\arraystretch}{1.1} 
    \caption{Accuracy measured on the test dataset}
    \centering
    \begin{tabular}{r | r r r}
        model \# & $r_{\mathrm{median}}$ & $r_{\mathrm{max}}$ & $r_{\mathrm{max}}$ [\%]\\
        \hline
        1 &  7.35e-06 & 2.40e-05 & 0.002 \\
        2 &  1.10e-05 & 5.43e-05 & 0.005 \\
        3 &  9.49e-05 & 5.22e-04 & 0.052 \\
        4 &  2.97e-04 & 1.45e-03 & 0.145 \\
    \end{tabular}
    \vspace{1mm}
    \begin{flushleft}
    \small
    \textit{Note --} Measurements are in terms of relative residual $r$, defined in Eq.~\eqref{eq:relres}, for the four neural networks analyzed.
    \end{flushleft}
    \label{tab:MBB_relres}
\end{table}

\subsection{Performance}\label{ssec:case1_perf}
A first benchmark concerns the time taken to process a batch of 10\,M samples\footnote{For reference, 10 - 100\,M is the typical order magnitude of the number of pixels of full sky maps treated within pipelines such as Commander.}, for different numbers of CPU cores.
All the measurements are taken through the method \texttt{adaptive\_autorange} exposed by the \texttt{benchmark} module of PyTorch.
It provides a robust adaptive routine, which takes care of the warm up and compilation cycles, as well as repeating the measurement until the variability, estimated as the inter-quartile range (IQR), falls below 0.05\% of the median, which is then taken as the measured value.

\begin{figure}[]
    \centering
    \includegraphics[width=.95\linewidth]{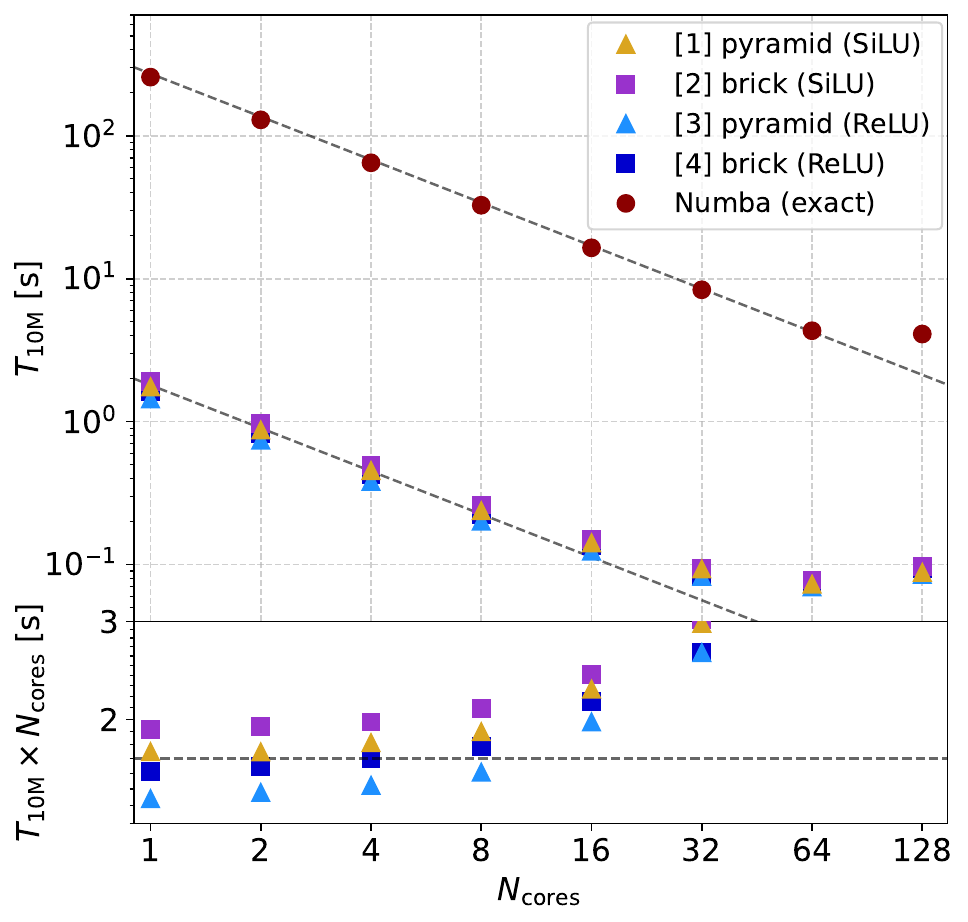}
    \caption{Benchmark results for the 4 network architectures listed in table~\ref{tab:MBB_4net}. 
    The top panel shows the actual timings compared to the ``brute force'' computation of the exact integral value, optimized and compiled through \texttt{Numba}, for a batch of 10\,M samples.
    The bottom panel shows the same run times multiplied by the corresponding number of cores employed, ignoring the exact integration results to better display the difference between the four models tested.
    In both panels, the dashed black lines show the slope of a perfect scaling. 
    All the axis scales are logarithmic, with the x-axis in base 2.}
    \label{fig:bench353}
    \vspace{20.pt}
\end{figure}

The benchmark has been performed on a cluster node featuring $4\times18$ 2.1-GHz cores (Intel E7-8870v3) cores, and the obtained run timings are shown in Fig.~\ref{fig:bench353} in comparison to the ``brute force'' exact integration, parallelized and compiled through \texttt{Numba}~\citep{Lam_2015_Numba}, used as a reference.

    The presented method appears to be faster by more than 2 orders of magnitude than an optimized direct summation routine. 
    In fact, for a bandpass profile of $\sim 1000$ points in frequency, the ``brute force'' summation requires approximately 10\,kFLOP per sample\footnote{The numbers reported count \texttt{exp} and \texttt{log} calls as a single FLOP, however, the real computational complexity of such transcendental operators strongly depends on the hardware and math libraries employed.}, versus the 300 - 400\,FLOP required by a forward pass on a neural network of this kind only.
    Moreover, \texttt{PyTorch} is natively optimized at a very low level through SIMD (Single Instruction, Multiple Data) techniques, such as AVX-512, and multi-threading, for efficiently processing large batches of data at once \footnote{
    Exploring the optimization of the brute force integration routine, beyond \texttt{Numba}'s capabilities, was outside the scope of this work. Therefore, the comparison shown here may not be an entirely fair competition between methods, but is surely a good reference point for the work presented.}.
    
    Among the four models tested, as expected from what is described in Subsect.~\ref{ssec:af}, the two featuring the SiLU as activation function recorded the slowest run times, by roughly a factor of 1.5. 
    However, the slowdown is compensated by a gain in terms of accuracy of more than one order of magnitude, as reported in Table~\ref{tab:MBB_relres}.
    On the other hand, the deeper structure of ``brick''-shaped networks requires to split the whole forward pass in one extra tensor product, yielding slightly slower inference time.
    
    The divergence of the recorded run times from an ideal scaling shows the wall set by the memory bandwidth limit. 
    For 128 cores, there is even an inversion in the scaling profile. 
    The number of physical cores of the machine (72, in this case) is exceeded, and concurrent threads trying to make use of the same CPU disrupt each other's activity, slowing down the whole computation. 
    This shows hyperthreading's limits in targeting well-optimized computations such as the tensor products driving a neural network.

Given the small size of the networks tested, and the fact that the first and most natural application of the method presented in this paper will be within Commander, which currently does not support any GPU computation for a series of reasons, the main interest is focused on optimal CPU performance.
However, to give an idea of the speed up that can be provided by a GPU, a second benchmark has been performed on a cluster node featuring 2$\times$NVIDIA P100 (with 3584 cores and 16\,GB of VRAM), for the same four models and the same 10\,M-samples batch.
The results are shown in Table~\ref{tab:bench_MBB_GPU}, in comparison with the fastest times recorded on the CPU, i.e. with the 64-cores setup.
The pure inference time on the GPU appeared to be around twice as fast in this case. 
However, the time taken to transfer the data to the GPU itself, needed if the neural network is employed in a CPU-based pipeline, has been timed at 15.4\,ms, reducing the gap between the two settings.

\begin{table}[]
    \setlength{\tabcolsep}{6pt} 
    \renewcommand{\arraystretch}{1.1} 
    \caption{GPU benchmark results compared to the 64-cores CPU setup}
    \centering
    \begin{tabular}{r | r r}
        model \# & 64-cores CPU wall time & GPU wall time \\
        \hline
        1 &  72.8\,ms & 31.6\,ms \\
        2 &  77.9\,ms & 36.2\,ms \\
        3 &  69.5\,ms & 31.5\,ms \\
        4 &  75.5\,ms & 36.3\,ms \\
    \end{tabular}
    \vspace{1mm}
    \begin{flushleft}
    \small
    \textit{Note --} The times reported are the medians measured through \texttt{adaptive\_autorange} from \texttt{PyTorch}, with a IQR-to-median ratio threshold set to 0.05\%, for the four different networks, on a 10\,M-samples batch.
    \end{flushleft}
    \label{tab:bench_MBB_GPU}
\end{table}

\section{A more advanced scenario: 9-parameters Astrodust+PAH}\label{sec:examp2}
\subsection{Setup}
As a demonstration of the capabilities of the method presented in this paper, a more complex scenario is now explored: a bandpass integration where the SED is given by the thermal emission of the Astrodust+PAH~\citep{Hensley_2023} model, from now on referred to simply as Astrodust.
The model features two main components: \textit{astrodust}, describing larger grains constituted by a single composite material, and various nanoparticle-sized grains such as the polycyclic aromatic hydrocarbons, or \textit{PAHs}.
Astrodust is able to describe precisely the interstellar dust emission (and extinction) through a physical description of the dust grains composition, size and temperature distributions, making it suitable for a variety of data analysis applications within cosmology and astrophysics.

As presented in~\citep{Hensley_2023}, Astrodust's thermal emissivity SED is given by
\begin{align}\label{eq:Ad_emiss}
\frac{I_\lambda^{\rm th}}{N_{\rm H}} = \sum\limits_{i = 1}^N \int da &\left(\frac{1}{n_{\rm H}}\frac{dn_i}{da}\right)\times \nonumber \\
&\int dT \left(\frac{dP}{dT}\right)_{i,a} C^{\rm abs}_{\rm ran}\left(\lambda, i, a\right) B_\lambda\left(T\right),
\end{align}
where $B_\lambda\left(T\right)$ is the Planck function, $(dP/dT)_{i,a}$ and $dn_i/da$ are the grain temperature and size distributions respectively, and $C^{\rm abs}_{\rm ran}\left(\lambda, i, a\right)$ denotes the extinction cross section for randomly oriented grains of composition $i$ and effective radius $a$, for radiation of wavelength $\lambda$.
$N_{\mathrm{H}}$ and $n_{\mathrm{H}}$ are the column\footnote{Integrated along the line of sight.} and volume number densities of hydrogen atoms, respectively.
For the purpose of the use-case here presented, we will keep the temperature distribution term unchanged and explore the dependency of the emissivity on the grain size distribution of the Astrodust component.
According to~\citep{Hensley_2023}, the latter is modeled as
\begin{align} \label{eq:dnda}
    \frac{1}{n_{\rm H}} &\frac{dn_{\rm Ad}}{da}\left(a\right) =  \frac{B_{\rm Ad}}{a} {\rm exp} \left\{ -\frac{ \left[\ln\left(a/a_{0,{\rm Ad}}\right)\right]^2}{2\sigma^2_{\rm Ad}} \right\} + \nonumber \\
    &\frac{A_0}{a}~{\rm exp}\left\{\sum_{i=1}^5 A_i \left[\ln\left(\frac{a}{\text{\AA}}\right)\right]^i\right\},
\end{align}
for 4.5\,\AA\ $< a < 5\,\mu$m, and set to zero otherwise.
Here $B_{\rm Ad}$, $a_{0,{\rm Ad}}$, $\sigma_{\rm Ad}$, and $A_i$ for $i \in [0,5]$ are scalar free parameters.

The functional form of the thermal emissivity is thus very complicated, and, as shown, is obtained through the integration of multiple non-trivial distributions.
The model has a total of 11 free parameters\footnote{If taking into account also the alignment function parameters, which only impact the polarized emission, not relevant for the case we are analyzing, the total number is 14.}, listed in Table~1 of~\citep{Hensley_2023} together with the fiducial values fitted by the authors during their analysis.
Fortunately, some Jupyter Notebooks have been made available, together with a series of fits files of pre-computed low-level variables, which allow to compute a functional form for the emission (and absorption) SEDs.
However, the code published is given as a demonstration to understand the model, rather than a high performance module ready to be integrated in an iterative analysis.
For this reason, its run times for producing an SED, especially when the grain size distribution parameters are changed, soon become prohibitive for a large-scale analysis through a MCMC-driven approach such as Commander.

The goal of test case presented here, is to exploit the neural network approach to predict the integral value given a set of 9 parameters, composed by:
\begin{itemize}
    \item 6 of the 11 low-level free parameters listed in Table~1 of~\citep{Hensley_2023}, namely $B_{\mathrm{Ad}}$, $a_{0,\mathrm{Ad}}$, $\sigma_{\mathrm{Ad}}$, $A_0$, $A_1$ and $A_2$ from~\eqref{eq:Ad_emiss}.
    \item A frequency-independent scaling factor regulating the energy density of the field heating the dust, namely $\log{U}$, defined in equation (9) of~\citep{Hensley_2023}.
    \item The dust surface mass density, integrated along the line of sight, named $\Sigma_{\mathrm{d}}$. It directly correlates with the emissivity as it is a multiple of the number density of hydrogen atoms $N_{\mathrm{H}}$, introduced in equation~\eqref{eq:Ad_emiss}.
    \item A parameter $q_{\mathrm{PAH}}$, regulating the relative abundance, and thus emission intensity, of the PAH with respect to the Astrodust component, where $q_{\mathrm{PAH}} = 1$ gives the default model.
\end{itemize}
Consequently, the parameter array $\mathbf{\Theta}$ now reads
\begin{equation}
    \mathbf{\Theta} = (B_{\mathrm{Ad}}, a_{0,\mathrm{Ad}}, \sigma_{\mathrm{Ad}}, A_0, A_1, A_2, \log{U}, \Sigma_{\mathrm{d}}, q_{\mathrm{PAH}}).
\end{equation}

\begin{figure}[]
    \centering
    \includegraphics[width=0.95\linewidth]{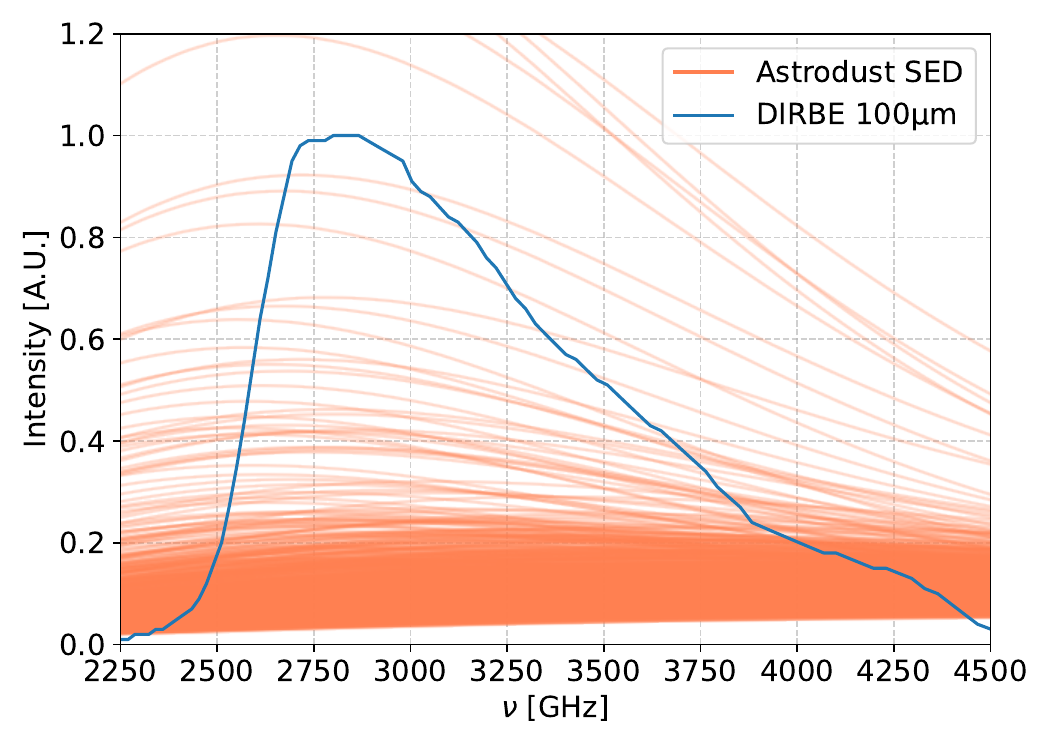}
    \caption{Dirbe 100$\,\mathrm{\mu}$m bandpass profile plotted together with a bundle of Astrudst SEDs generated for random realizations of the parameters within the explored region. Note that the y-axis is represented in arbitrary units, as multiplicative factors, such as unit conversions, will not affect the analysis presented.}
    \label{fig:case_2_bp}
    \vspace{12.pt}
\end{figure}

Since the dust emission is dominant over other foregrounds within the infrared region of the spectrum, the SED is corrected through the bandpass profile of COBE Dirbe~\citep{mather:1994} 100$\,\mathrm{\mu}$m instrument, which is shown in Figure~\ref{fig:case_2_bp}.

\subsection{Training}\label{ssec:case2_training}
A range of approximately $\pm10\,\%$ of the best fit values reported in~\citep{Hensley_2023} is again chosen as the training region for the 9 free parameters, as listed in Table~\ref{tab:paramsAd}.
\begin{table}[]
    \setlength{\tabcolsep}{6pt} 
    \renewcommand{\arraystretch}{1.1} 
    \caption{Training ranges for the 9 free parameters}
    \centering
    \begin{tabular}{r | r r r}
        Parameter & Fiducial & Train Lo & Train Hi \\
        \hline
        $B_{\mathrm{Ad}} \times 10^{10}$ &  3.31 & 2.65 & 3.97 \\
        $a_{0,\mathrm{Ad}}$\,[Å]&  63.8 & 51.04 & 76.56 \\
        $\sigma_{\mathrm{Ad}}$ & 0.35 & 0.28 & 0.42\\
        $A_0\times10^{5}$ & $2.97$ & $2.38$ & $3.56$\\
        $A_1$  & -3.40 & -4.08 & -3.30 \\
        $A_2$  & -0.81 & -0.97 & -0.78\\
        $\log{U}$ & 0.20 & 0.10 & 0.30\\
        $\Sigma_{\mathrm{d}}$ & ... & 1.00 & 2.00 \\
        $q_{\mathrm{PAH}}$ & 1.00 & 0.90 & 1.10\\
    \end{tabular}
    \vspace{1mm}
    \begin{flushleft}
    \small
    \textit{Note --} When applicable, the fiducial value found in~\citep{Hensley_2023} is reported for each parameter.
    \end{flushleft}
    \label{tab:paramsAd}
\end{table}
This time, to obtain an efficient sampling of the 9-dimensional hypercube without exponentially increasing the total number of samples, a quasi-Monte Carlo method, namely a Sobol sampler~\citep{sobol1967distribution}, is employed.
Instead of using purely random numbers, which in high dimensions tend to cluster in clumps and leave large empty spaces, a Sobol sequence places points in a deterministic pattern designed to evenly cover the target space~\citep{Sobolcoverage}.
This way, a chain of $2^{23} \approx 8\,\mathrm{M}$ samples\footnote{The powers of 2 are the optimal number of elements for an uninterrupted Sobol sequence.} is produced.
As usual, the exact integral value is computed for each of them, yielding the targets.
The full dataset produced, where each value has been stored as double-precision float, is around 640MB in size.
Fig.~\ref{fig:case_2_bp} shows an ensemble of Astrodust SEDs produced from a subset of $2^{10}$ random samples of the parameters, produced within the same boundaries as the training dataset, to give and idea of the higher variability of the targeted SEDs, if compared to the previous test case (Fig.~\ref{fig:case_1_bp}).

Similarly as in Sect.~\ref{sec:examp1}, four neural networks are compared, of different shapes and number of trainable parameters, listed in Table~\ref{tab:Ad_4net}. This time, a SiLU activation function is employed in all models, as it has already been proven to provide better prediction accuracy than the default choice ReLU.

\begin{table}[ht]
    \setlength{\tabcolsep}{6pt} 
    \renewcommand{\arraystretch}{1.1} 
    \caption{Main features of the 4 neural networks compared.}
    \centering
    \begin{tabular}{r | c r}
        model \#  &  Shape &  $N_{\mathrm{t}}$ \\
        \hline
        1 & [9,16,16,16,16,1] & 993 \\
        2 & [9,32,32,32,1] & 2465\\
        3 & [9,64,32,1] & 2753\\
        4 & [9,64,32,16,1] & 3265\\
    \end{tabular}
    \vspace{1mm}
    \begin{flushleft}
    \small
    \textit{Note --} The first column gives an arbitrary indexing, useful for referring to plots in Figures~\ref{fig:training_case2} and~\ref{fig:histogram_case2}. The network shape is given in the second column as an array, where each element gives the number of nodes of the corresponding layer, while the last column gives the number of trainable parameters as a measure of computational complexity. The actual total number of FLOPs can be estimated through the rule of thumb $N_{\mathrm{FLOPs}} \approx 2 \times N_{\mathrm{trainable}}$, even though the operations given by a complex activation function such as the SiLU, in common between the four networks, should be counted as well.
    \end{flushleft}
    \label{tab:Ad_4net}
\end{table}

\begin{figure}[]
    \centering
    \includegraphics[width=0.95\linewidth]{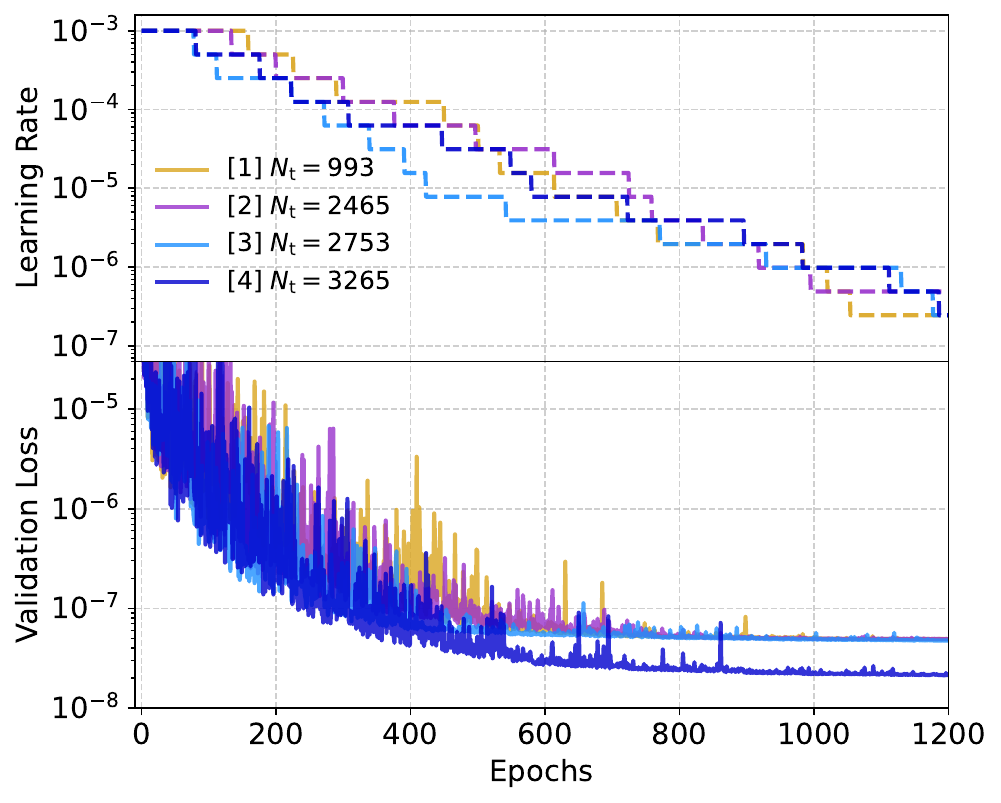}
    \caption{Training metrics of the four network architectures listed in Table~\ref{tab:Ad_4net}. The dashed lines in the top panel show the learning rate decay, dictated by the scheduler.
    The bottom panel shows the averaged validation loss function for each network, which is correlated to the residuals through Eq.~\eqref{eq:Huber}. }
    \label{fig:training_case2}
    \vspace{10.pt}
\end{figure}

The four networks are then trained with a very similar routine to the one described in Subsect.~\ref{ssec:case1_training}, with the exception that, given the bigger size of the networks, the training batch size has been increased to 4096 and the initial learning rate to $10^{-3}$. 
In order to speed up the whole process, the training is this time performed on a GPU Nvidia P100, which is well capable of storing each network and the dataset simultaneously within its 16GB of VRAM. 
With the setup described, the training takes around 12s per epoch.
In this case, however, as shown in Fig.~\ref{fig:training_case2},  around 1000 epochs are required to see the loss functions plateauing at their final values, making the whole process several times more time-consuming than the previous example described in Sect.~\ref{sec:examp1}.
Given the bigger size of the networks trained, the landscape of the the loss functions is much more complex, yielding noisier trace plots until the learning rate has not decayed.

\subsection{Accuracy}
Similarly to the previous test case, for assessing the prediction accuracy of the trained networks, a test dataset composed by $2^{20} \approx 1\,\mathrm{M}$ samples is generated with the same technique described in Subsect.~\ref{ssec:case2_training}, within boundaries roughly 10\% narrower than the training dataset.
The same relative residual, defined in Eq.~\eqref{eq:relres}, is used as a metric for assessing accuracy.
This time, however, the high number of dimensions makes it impossible to visually explore the error's behavior within 2-dimensional slices of the parameter space, as in Fig.~\ref{fig:valid_multifig}.
Instead, the percentage distribution of the relative residuals $r$ on the test dataset is computed, and plotted as a histogram in Fig.~\ref{fig:training_case2}.
As expected, increasing the network complexity squeezes the corresponding residual distributions towards zero.
In all the four cases we measure that 99.9\% of the relative errors are well below 0.1\%, taken as a general reference.
In fact, given the high complexity of the residual landscape in a 9-dimension space, the $99.9^{\mathrm{th}}$ percentile gives a more robust single-valued estimate of the general prediction accuracy of the networks than the maximum relative residual, which can be dragged away by a few isolated outliers.
For reference, the network \#4 measured a maximum relative error of 0.08\%.

\begin{figure}[]
    \centering
    \includegraphics[width=0.95\linewidth]{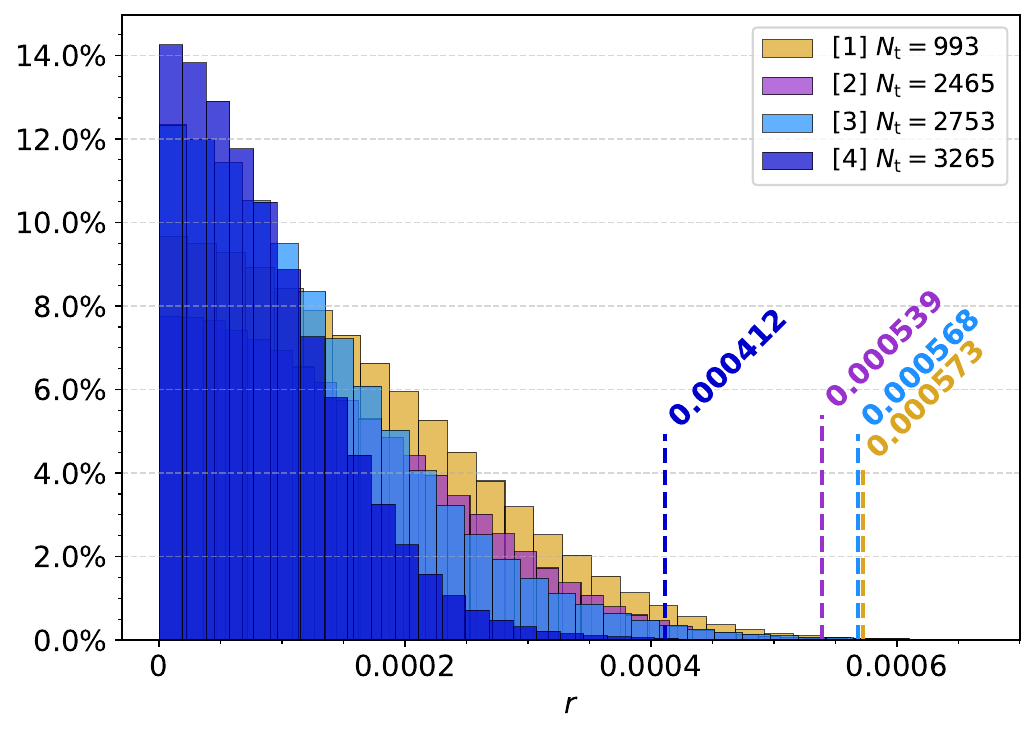}
    \caption{Distribution of the relative residual for the four neural networks trained. The frequency of each bin is expressed as a percentage of the total number of test samples ($2^{20}$).
    The vertical dashed lines indicate the position of the $99.9^{\mathrm{th}}$ percentile.}
    \label{fig:histogram_case2}
    \vspace{10.pt}
\end{figure}

\subsection{Performance}
\begin{figure}[]
    \centering
    \includegraphics[width=0.95\linewidth]{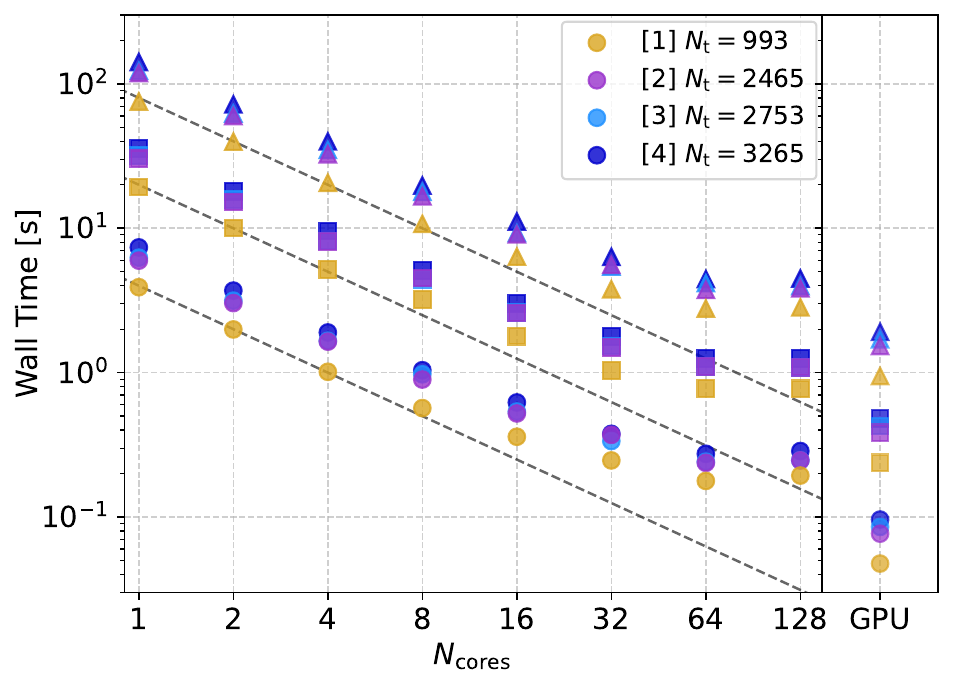}
    \caption{The left panel shows the benchmark results for the 4 network architectures listed in table~\ref{tab:Ad_4net}, obtained for the three batch sizes of 10\,M (circular markers), 50\,M (square markers) and 200\,M (triangular markers) respectively. While the dashed black lines show the slope of a perfect scaling.
    The right panel shows the same measurements obtained on the Nvidia P100 GPU. All the axis are represented in logarithmic scale, with the x-axis in base 2.}
    \label{fig:benchAd}
    \vspace{12.pt}
\end{figure}

A benchmark of the integral computation times obtained by the neural networks against the straightforward evaluation and integration of Astrodust+PAH SED would not be a totally fair comparison.
In fact, as already mentioned, the code is given as a tool to explore the model and use it statically, rather than an optimized computational method for iterative data analysis.
Its evaluation takes several orders of magnitude longer than the neural networks employed here.
For this reason, only the measured run times for the four neural networks are taken into account, given different number of CPU cores and sample batch sizes.
The inference is timed on the same node, featuring $4\times18$ 2.1-GHz cores (Intel E7-8870v3), for batches of 10, 50 and 200 millions of samples respectively, roughly corresponding to the number of pixels in full-sky HEALPix~\citep{Gorski_2005} maps with a resolution parameter $N_{\mathrm{side}}$ of 1024, 2048 and 4096 respectively.
The results are shown in Fig.~\ref{fig:benchAd}, where the run times measured on the same GPU Nvidia P100 are included as well.
The scaling behavior is very similar to what has been observed in the previous test case, described in Subsect.~\ref{ssec:case1_perf}.
This time, however, the usage of a GPU provided roughly a $2-3\times$ speed up, compared to the fastest CPU run time of each benchmark, which again resulted being for 64 cores.
As expected, the usage of a GPU provides a bigger speed up as the problem size increases.

\section{Conclusions}
This work presented a new method for approximating the smooth function of the SED's parameters, giving the bandpass integral correction, through a neural network.
The method has been applied to both a simpler 3-parameter test case, with a modified black body as SED and a bandpass from Planck HFI 353\,GHz, and a more complex 9-parameter one, based on the expensive Astrodust+PAH emission model.
In both cases four network architectures with different sizes, shapes and activation functions have been analyzed.
The SiLU activation function resulted being the optimal choice for the purpose of the method presented.
In both test cases at least one or more of the models tested provided a satisfactory result in terms of accuracy, with maximum relative residual as low as 0.002\% and 0.08\%, respectively.
The high accuracy obtained came together with a significant save in computation run time, with speed ups of more than two orders of magnitude compared to an optimized Numba-compiled version of the exact integral summation routine, as shown by the first test case.
The method developed and presented in this paper has been published in the form of a new Python package, \texttt{pyfine}, making it easily accessible and usable.

\section*{Acknowledgments}

The current work has received funding from the European Union’s Horizon 2020 research and innovation programme under grant agreement numbers 101141621 (ERC; \textsc{Commander}). This article reflects the view of the author only. 
The funding body is not responsible for any use that may be made of the information contained therein. 
The work presented made use of data stored in the Legacy Archive for Microwave Background Data Analysis (LAMBDA), part of the High Energy Astrophysics Science Archive Center (HEASARC). HEASARC/LAMBDA is a service of the Astrophysics Science Division at the NASA Goddard Space Flight Center.

Most of the work presented in this paper has been carried out in the context of the PhD course FYS9429 ``Advanced machine learning and data analysis for the physical sciences'', held at the University of Oslo by professor Morten Hjorth-Jensen, whose expertise and guidance are acknowledged by the author.

Important help, support and exchange of ideas came from Hans Kristian K. Eriksen, Mathew Galloway, Jonas G. S. Lunde, Sigurd K. Næss, Duncan J. Watts, and the \textsc{Cosmoglobe} collaboration in general, throughout the development of the work here presented.
Finally, the author acknowledges the support received by Brandon S. Hensley with the theoretical understanding and usage of Astrodust+PAH model.

LLM tools have only been used during the preliminary phases of the work here presented, and only as an efficient research tool and for debugging purposes. No generative AI, LLMs, or AI-assisted writing tools were used in the drafting, conceptualization, or editing of this manuscript.

\bibliographystyle{apsrev4-1}

\bibliography{bpnn}

\end{document}